\documentstyle[preprint,aps,floats,epsf]{revtex}
\firstfigfalse

\def\mkfig#1{}

\def\mkfig#1{#1}

\def\lab#1{\label{#1} }  \def\bibit#1{\bibitem{#1} } \def\cit#1{\cite{#1}}

\def\beq{\begin{equation}}
\def\eeq{\end{equation}}
\def\bea{\begin{eqnarray}}

\def\CG{{\cal G}}
\def\CH{{\cal H}}
\def\CD{{\cal D}}
\def\li{\left\langle}
\def\re{\right\rangle}
\def\de{\delta}
\def\del{\nabla}
\def\0{\over}
\def\9{\partial}
\def\tr{\mbox{Tr} }
\def\om{\omega}
\def\eps{\epsilon }
\def\al{\alpha}
\def\be{\beta}
\def\lap{\triangle}
\def\bp{ p}
\def\bq{ q}
\def\lb{\left[ }
\def\rb{\right] }
\def\1{\vec}
\def\({\left(}
\def\){\right)}
\def\k{\kappa}

\def\qd{\quad }
\def\P{\mbox{Phys} }
\def\half{ {}^1 \hspace*{-0.2em} /_2 }

\def\ah{\hat A}
\def\et{E^T}

\begin{document}

\draft

\preprint{{\small E}N{\large S}{\Large L}{\large A}P{\small P}-A-510/95,
hep-ph/9503244 }

\title{\vspace*{\fill}
Finite temperature formalism for nonabelian gauge theories in the
physical phase space }

\author{Herbert Nachbagauer\footnote[3]{e-mail: herby @ lapphp1.in2p3.fr} }

\address{  Laboratoire de Physique Th\'eorique ENSLAPP
\footnote{URA 14-36 du CNRS, associ\'ee \`a l'E.N.S. de Lyon,
et au L.A.P.P. (IN2P3-CNRS)\\
\hspace*{0.7cm} d'Annecy-le-Vieux} \\
Chemin de Bellevue, BP 110, F - 74941 Annecy-le-Vieux Cedex,
France}

\date{\today}

\maketitle

\begin{abstract}
We establish a new framework of finite temperature field theory for
Yang-Mills theories in the physical phase space eliminating all
unphysical degrees of freedoms.
Relating our method to the
imaginary time formalism of James and Landshoff in temporal
axial gauge, we calculate the two-loop pressure and provide
a systematic and unique method to construct the additional
vertices encountered in their approach.

\end{abstract}


\newpage

\section*{Introduction }

This paper is devoted to establish  finite temperature field theory
on the physical phase space of nonabelian gauge theories.
It is the intrinsic nature of every gauge theory that the whole
configuration space  contains gauge group orbits, and gauge
transformations generate shifts along those orbits.
Gauge equivalent field
configurations are physically indistinguishable, therefore only
transitions between distinct gauge orbits contain physical
information.

In standard field theory the problem of superficial degrees of freedom
is attacked by the introduction of a gauge fixing condition.
However,  due to Gribov's ambiguity
\cit{grib1}, the gauge orbit space of Yang-Mills potentials cannot
be parametrized uniquely by potentials satisfying a local gauge condition.
A gauge condition surface in the entire configuration space contains
gauge equivalent field configurations.

Moreover, there
are may exist certain field  configurations where the gauge
fixing surface is tangential to gauge orbits, corresponding
to the zeros of the Faddeev-Popov determinant. Perturbatively,
this entails,
depending on the gauge chosen, unphysical poles in the propagator that
have to be defined properly.
In particular for the class of axial gauges, $n\cdot A=0,\qd n^{\mu} =
(1,\1 n)$, the spurious poles at $p_0=\1p \1n$ have to  be
treated with the co-called Leibbrandt-Mandelstam \cit{LM} prescription,
which has to be modified \cit{gaig1} for $n^{\mu}$ lightlike.
It should be mentioned that Landshoff's $\al$-prescription
\cit{land0} also gives the correct exponentiation in a Wilson-loop
calculation up to order $g^4$.

At finite temperature in the real time formalism (RTF),
analogously to the zero temperature case,
a temperature-dependent pole-prescription
arises  naturally within the framework of Hamiltonian quantization
\cit{nach1}, and for the particular choice of temporal axial gauge (TAG),
($\vec n =0$), a RTF has been developed successfully \cit{jame0}.

In  the so-called imaginary time formalism (ITF),
however, the energy can only take on the discrete
Matsubara frequencies $p_0=2 \pi i n T$. Thus if one naively
heats the unphysical degrees of freedom,
an unresolved pole remains for the zero mode for momenta  $\1p \1n=0$.
Luckily, in the particular case of the unresummed
imaginary part of the transversal structure function of the gluon
self-energy, these factors cancel out for symmetry reasons
and the straight-forward application of ITF Feynman rules works \cit{nach2}.

At the contrary, in TAG the naive ITF propagator contains unregularized
singular factors $1/p_0$ at zero Matsubara frequency.
This problem has  been circumvented in earlier works \cit{kaja1}
by the ad hoc assumption that such poles have to be dropped.
Although there exists no justification for doing so, the leading order
self-energy is found to coincide with the results in
other gauges. The deeper reason for this is, however, not the correctness
of this prescription. In fact, in general axial gauge,
the dependence on the gauge fixing vector $\vec n$, and thus the axial
poles,  completely cancels  out algebraically \cit{nach2}
which gives the proof  that no prescription enters at that loop level.
In fact, a closer inspection of the
corresponding expressions in TAG reveals that this
cancellation takes place in that gauge too.
Moreover, this must be the case since
the leading-order expression for the
self-energy is nothing but the hard thermal loop, and thus a physical,
gauge independent quantity.

It is well known, that the consistent calculation of the next-to-leading
order contributions
requires an appropriate resummation of propagators and vertices  \cite{resum}.
However, since the two-point function is a gauge independent quantity
only on the mass-shell, one cannot expect the self-energy to be
prescription independent off the physical dispersion relation.
In the light of this line of arguments, the non-Debye screening
behavior \cit{peig1,baie1} which contradicts results obtained in
Coulomb gauge \cit{rebh2}, and using  Polyakov-loop correlators
\cit{rebh2,braa1} appears to be rather an artifact
of the off-shell calculation and not
of the ad hoc pole prescription. Adopting the on-shell definition of
the  Debye-mass proposed by Rebhan \cit{rebh1} we expect that in a
pragmatic calculation keeping the undefined $1/p_0$ quantities
those unphysical poles  cancel out.

Taking seriously the naive formulation, it was proposed quite recently
\cit{kala1} to regularize the divergent $1/p_0$  expressions by a temperature
dependent expression. However, this proposal may only serve to
give the expressions an intermediate meaning, and it has to turn out
irrelevant in the calculation of physical quantities. Moreover,
it is not clear
if this prescription leads a truly temporal propagator and
can be adopted unambiguously without  introducing ghost fields.
Independent of any ad hoc method to get rid of the temporal pole,
there  is no justification for the naive application
of ITF Feynman rules in TAG.

It has  been pointed out by James and Landshoff \cit{jame1} already
some time ago that the temporal pole
is related to the free motion of the longitudinal modes of the gauge
field. This violates periodic boundary conditions that are
necessary to set  up ITF. James and Landshoff invented a new formalism in which
the longitudinal fields  remain unheated and only the remaining
physical degrees of freedom attain a temperature. Within this
formalism, the longitudinal part of the propagator is automatically
free from the $1/p_0$ singularity. It was argued that one can obtain
the same answer for the two-loop pressure as in other gauges.
The main drawback in that formulation is, however, that one has to construct
physical states by explicitly solving the Gauss law which gives
rise to additional time-independent vertices.
This results in unwieldy expressions
and it appears  difficult to establish a resummation program.

In the present paper, we advocate a different route to attack the problem.
Based on a Hamiltonian formulation of the theory, we are able to
eliminate all unphysical degrees of freedoms from the Hamiltonian
by introducing an appropriate coordinate system in the space of field
configurations which allows to
make a unique distinction between gauge degrees of freedoms and
physical ones \cit{shab1}. Within this approach one
neither encounters any unphysical poles nor any explicit construction
of physical states is necessary. It is straightforward to heat the
physical degrees of freedom in the resulting non-local Hamiltonian.
In order to illustrate the new method,
we calculate the two-loop pressure and compare the result with
the corresponding expressions found using the approach of James and Landshoff.
Rewriting the physical fields in the basis used in
their investigation,  we give a general strategy to construct the
corresponding Gauss law states to arbitrary order  and
calculate them explicitly  to third order in the coupling constant.

\section{Construction of the Hamiltonian  in the physical phase space}

In this section we construct the physical Hamiltonian of pure QCD
by eliminating all gauge degrees of freedom. The
basic  idea may be illustrated as follows. Consider a point particle moving
in a plane and rotations around the origin as symmetry group. Then
different trajectories are gauge equivalent if they can be mapped one
to another by rotations. Of course, the dynamics appears
 simplest in
polar coordinates, where only the radial coordinate has a physical
meaning.  The angular momentum generates gauge transformations and the
corresponding canonical conjugate position coordinate, the angle, may be
fixed arbitrarily.

We start our investigation with the pure QCD Lagrangian
\beq L = -{1\0 2} \int d^3 x \,\tr \, F^{\mu\nu} F_{\mu\nu} = -{1\0 2}
\left< F_{\mu\nu}, F^{\mu\nu} \right> \lab{lag}
\eeq
where $F_{\mu\nu} = \9_\mu A_\nu - \9_\nu A_\mu + i g \left[ A_\mu ,
A_\nu \right],$ and the Yang-Mills fields are elements of the Lie
algebra of the gauge group\footnote{The hermitian generators of the
 gauge group are normalized according to $\tr \, T^a T^b = \de^{ab}/ 2$.}.
 To go over to the Hamiltonian  formalism,
we have to determine the canonical momenta $E^\mu = \de L/\de \dot
A^\mu=F^{0\mu} .$ The momentum conjugated to $A_0$ vanishes, $E_0 \sim 0 $
forming
the primary constraint. The corresponding canonical Hamiltonian has
the form $H=2 \li \dot A^i , E^i  \re - L =  \li E^i, E^i \re + V(A^i)
- 2 \li A_0 , \CG \re $ where
\beq V(A)=  {1 \0 2} \li F^{ij}, F^{ij}
\re \eeq
appears as potential and $\CG= \del^i[A] \; E^i = \9^i E^i - i g [A^i,E^i] $
is the so-called Gauss law,  $\del^i[A]$ being the  covariant
derivative in the  adjoint representation.

The primary constraint must be conserved during time evolution. This
yields the secondary constraints
$ \dot E_0 = \left\{ E_0 , H \right\} = \CG \sim 0 ,$
where we implicitly assumed the standard equal-time Poisson brackets
$ \left. \left\{  A_{\mu}^a (x)  , E_{\nu}^b (y) \right\}\right|_{x_0=y_0} =
\de^{ab} g_{\mu \nu}\de^3 ( x- y).$

Since the algebra of the Gauss law closes,
$\left\{ \CG^a (x) ,\CG^b (y)  \right\}  =
g f^{abc} \de^3 (x-y) \CG^c (x) $ and
$\dot \CG^a =
\left\{ \CG^a , H \right\} = - f^{abc} A_0^b \CG^c $
we conclude that there are no more
constraints, and all constraints are of the first class.
Following the standard Dirac quantization procedure \cit{dira}, the Poisson
brackets may be deformed to eliminate the constraints $E_0 \sim 0$
at the operator
level. Since the gauge orbits generated by  $E_0$ are  shifts of $A_0$ only,
and leave the other phase space variables $A^i, E^{\mu} $ untouched,
we select the gauge equivalent configuration in the phase space which
satisfies $A_0=0$. This amounts to simply dropping the canonical
pair $A_0,\; E_0 $ from the Hamiltonian.
On this hyperplane, the remaining constraints $\CG$ become  time
independent since the Hamiltonian now commutes
with the Gauss law. Those constraints  generate time
independent gauge transformations on the remaining phase space
variables $E^i$ and $A^i ,$
$$ E^{\Omega} = \Omega E \Omega^{-1} , \qquad A^{\Omega} = \Omega A
\Omega^{-1} -{i \0 g} ( \9 \Omega ) \Omega^{-1} ,$$
where $\Omega$ is an element of the gauge group.
It is  convenient to formulate the quantized theory in a functional
representation.  Representing the canonical momenta by the standard
functional differential operator $ E (x) \to -i \de /  \de A (x) $ the
Schr{\"o}dinger equation takes the form
\beq H \Psi_n [A] =
\left( -  \li {\de \0 \de A } , {\de \0 \de A } \re +  V[A] \right)
\Psi_n [A] = E_n \Psi_n [A] \lab{sch} \eeq
and the wave function is subject  to the constraints
\beq \CG \Psi_n [A] = \del[A] {\de \0 i\de A } \Psi_n [A]  = 0 .
\lab{gl} \eeq
In the approach of James and Landshoff \cit{jame1} the  wave
functions are  constructed explicitly by solving this constraint
in  order to be able to
perform the  thermal trace over physical states. Alternatively, one
may eliminate the superficial degrees of freedom by reducing the
number of field components. In analogy to  the example given in the
beginning of the section, we parametrize the  field
configurations in the unconstraint configuration space
by an 'angle' $\om$ and the
remaining coordinates $ \hat A{}^i$ in the following way
\beq  A = U \hat A U^{-1} - {i \0 g } ( \9 U ) U^{-1} .\lab{tra}
\eeq
Here $U[\om] = \exp{( i g \om )} $ is an element of the gauge group
generated by the Lie-algebra valued angle $\om$ and $\hat A^i = \eps^i_{\al}
A^{\al} , \,  \al=1,2$ are the remaining coordinates projected out by the
operator
$\eps$ that is normalized such that $\eps^i_{\al} \eps^i_{\be} =\de_{\al \be} $
acts as unity in the corresponding  subspace.
The constraint (\ref{gl}) transforms into
\beq \CG \; \Psi [\hat A ,\om ] = U {\de \0 \de
 i \om }  \Psi[\hat A, \om ]  U^{-1}  =0  \lab{gl2} \eeq
which tells us that physical variables are independent of the angle
$\om$. The Gauss law itself turns out to be the canonical conjugate
momentum to the 'position' variable $\om$.

In the potential part $V[A]$ we may simply replace the original gauge
field by its
physical components, $V[A] = V[\hat A]$, since it is gauge invariant under
the transformation (\ref{tra}) . For the
kinetic part in the Hamiltonian, we need an expression for metric
entering implicitly in the definition of the inner
product  in (\ref{lag}). The lower metric components can be
read off from the  differentials, (The
anihermitian  covariant derivative is meant to be taken with
respect to the fields $\hat A$.)
$$ \li \de A , \de A \re = \li \de {\hat A}^{\dag}  , \de \hat A \re +
\li  \de \om \del^{\dag} , \de \hat A \re
+ \li \de \hat A^{\dag} ,
\del \de \om \re + \li \de \om , \del^{\dag} \del \de \om \re $$
i.e. $g_{\al \be} =  {\eps^i_\al}^{\dag} \eps^i_{\be} = \de_{\al \be} ,
\quad   g_{3 \al} =g_{\al 3}^{\dag}  =   \hat \del_{\al} ,
\quad g_{33}  =
{\del^i}^{\dag} \del^i $,  and we defined the projected covariant
derivative as $\hat \del_{\al} = \del^i \eps^i_{\al} $.
We shall also need the determinant of the metric, $\mu^2 = \det (g)=
\det (\CD),
\quad \CD = \del^{\dag}  \del - \hat \del^{\dag} \hat \del $,
and the inverse components,  which read
$g^{\al \be}=\de^{\al \be} + {\hat\del{}^{\al}}^{\dag} {\CD}^{-1}
{\hat\del}^{\be},
\quad g^{\al 3}= {g^{3 \al}}^{\dag} = - {\hat \del{}^{\al}}^{\dag} {\CD}^{-1} ,
\quad g^{33} ={\CD}^{-1} .$
The kinetic term in the Hamiltonian reads in
covariant form
$ \li \de /\de X^i , \de / \de X^i \re  = \li \mu^{-1} \de / \de X^A \;
g^{AB} ,  \mu \; \de  / \de X^B \re $
where we have put $X^A = ( A^{\al},\om ).$
When this operator acts on   physical wave functions which by virtue of
(\ref{gl2}) do not
depend on the angle $\om$, the terms containing the  momenta $\de / \de\om$
vanish. Thus the physical Hamiltonian is obtained by simply dropping
the 3-components in the metric and the Schr{\"o}dinger equation
(\ref{sch}) together with the constraint (\ref{gl}) is equivalent
to the reduced dynamical system  described by the Hamiltonian
\beq H^{phys}   =  - \li  \mu^{-1}
\left ({\de \0 \de A^{\al} }\right)^{\dag} ,
( \de^{\al \be} + {\hat \del{}^{\al}}^{\dag} {\CD}^{-1} {\hat \del{}^{\be} }
 ) \mu {\de \0 \de A^{\be} } \re
+ V[\hat A]   \lab{ham}  \eeq
where the wave function as well as observables are functionals of
the coordinates $\hat A$ only. Apart from the determinant
$\mu$ appearing in the Hamiltonian, this   expression
was already found in \cit{creu} using a different construction. The
determinant resolves the apparently existing operator ordering problem
which is due to the necessary  inversion of the operator $\CD$ in the
kinetic term. This problem was also discussed previously \cit{turd},
however without definite solution.  The
operator $\mu$ gives contributions $\sim (\de^3(0))^2$, which may be dropped
if one is  only interested in the local properties of the theory.

\section{Feynman rules in the physical subspace}

The Hamiltonian derived in the previous section may serve to establish
a set of Feynman rules in the physical subspace. Since it no more
contains unphysical degrees of freedoms, it is straightforward to
establish finite temperature field theory by heating the field $\hat
A$.
We emphasize that no explicit choice of the projection operator $\eps^i_{\al}
$ corresponding to  particular coordinates is necessary
so far.  However, (\ref{ham}) contains a non-local
operator which makes the theory unwieldy to deal with.
Alternatively, one may introduce an auxiliary  field and
rewrite the  Hamiltonian density in the following manner
\beq {\cal H} = {1 \0 2} { E^{\al ,a}}^{\dag }  E^{\al,a}  +
{1 \0 2 } \left[ {\Phi^a}^{\dag}
({\hat \del}^{\al}  E^{\al} )^a + {( \hat \del{}^\al E^\al )^a}^{\dag}
\Phi^a \right] - {1 \0 2 } {\Phi^a}^{\dag} {\CD }^{ab} \Phi^b + {1 \0 4}
{\hat F}^{ij,a} {\hat F}^{ij,a} + {1 \0 2 } {\rho^a}^{\dag}
{\CD}^{ab} \rho^b . \lab{ham3} \eeq
The electric field $\hat E$  is the canonical conjugate to
$\hat A$ in operator representation, $\left. [{ E}(x)^{\al,a} ,
 { A}(y)^{\be,b}] \right|_{x_0 = y_0} = -i\de^{ab}\de^{\al\be}
\de^3(x-y) $.
The last term in (\ref{ham3}) just subtracts off the trace
of the operator $\CD$,
which amounts to drop all Feynman graphs which do not contain at least
one $ \Phi^a ({\hat \del}^{\al} { E}^{\al} )^a$
vertex.

For the particular choice of purely
transversal fields, $ \9^i {\hat A}^i=0$, the operator sandwiched between the
covariant derivative and  the electric field
becomes the spatial transversal projection operator,
$\eps^i_{\al} \eps^j_{\al} = \de^{ij} - \9^i \9^j/ \9^2 = T^{ij}(\9) $.
In that case,
the $\Phi \del E$ terms in (\ref{ham3}) turn into a single three-vertex
$\Phi \hat A \hat E$ and not two-vertex $\Phi \hat E$ remains.
We observe that even for a non-transversal choice of $\hat A, \hat E$,
the perturbative ${\hat E}^L {\hat E}^L$ propagator is compensated by the
${\hat E}^L \Phi$ two-vertex  and the $\Phi \Phi $ Green function. This
corresponds to the fact that in pure QCD  zeroth order longitudinal
states are pure gauge degrees of freedoms. We want to point out
that although all
fields are purely transversal they do not coincide with the
transversal components of the original gauge potentials, and
the transversal choice must not be confused with the Coulomb gauge.

Splitting the  Hamiltonian into a free part
$$ {\cal H_0 } = {1 \0 2 }
\left( E^{a,i} T^{ij} E^{a,i} - A^{i,a} \lap T^{ij} A^{i,a} + \Phi^a
\lap \Phi^a \right) $$
and an interacting one
\begin{eqnarray*}  \CH_{WW} &= &\CH^{\Phi A E} +
\CH^{\Phi A} +  \CH^{V(A)} ,\\
\CH^{\Phi A E} &= & g f^{abc} \Phi^a E^T{}^{i,b}  A^T{}^{i,c} ,\\
\CH^{\Phi A} & = & g f^{abc}\Phi^a A^T{}^{i,c} \9^i \Phi^b
-{g^2 \0 2 } f^{abc} f^{ade} \Phi^b A^T{}^{i,c} A^T{}^{j,e} {\9^i \9^j \0
\9^2} \Phi^d ,\\
\CH^{V(A)}&=&- g f^{abc} \9^j A^T{}^{i,a} A^T{}^{j,b} A^T{}^{i,c} +
{g^2 \0 4} f^{abc} f^{ade} A^T{}^{i,b} A^T{}^{j,c} A^T{}^{i,d} A^T{}^{j,e},
\end{eqnarray*}
the former one gives rise to the propagators
 \unitlength1cm
\begin{eqnarray*}
\parbox{4cm}{
\begin{picture}(0,0)
\includegraphics{prop1.eps}
\end{picture}   }
 = & { \li A^{Ti} A^{Tj} \re}(p_0,{\1p } )& =
- T^{ij}({\1p} )
{1 \0 p_0^2 - \vec p^{\, 2} }\\
\parbox{4cm}{ \vspace*{1.5em}
\begin{picture}(0,0)
\includegraphics{prop2.eps}
\end{picture}   }
 = &  { \li E^T{}^i E^T{}^j \re}(p_0,\1p) & =
- T^{ij}(\1p) { p^2 \0 p_0^2 - \vec p^{\, 2 }} \\
\parbox{4cm}{
\begin{picture}(0,0)
\includegraphics{prop3.eps}
\end{picture}   }
 = & { \li \Phi \Phi \re }(p_0,\1p) & = - {1 \0 \1p^{\, 2 } },\end{eqnarray*}
and from the second one we shall only need the expression for the
$\Phi^a A^T{}^{i,b} E^T{}^{j,c}$ vertex given by
$ (- 1) g f^{abc} \de^{ij} . $

For two reasons, we do not express the electric fields in terms of
time derivatives of the
gauge potentials. Firstly, this would again involve inverting non-local
operators, and subsequently render the theory untractable. Secondly,
the kinetic term and the potential term are both by themselves
physical observables. Keeping the $E$ fields, one preserves the possibility to
calculate the electric and magnetic dispersion independently.

At the present stage, the propagator of the field $\Phi$ only serves
to write the inverse Laplacian in a way
convenient for calculating quantities in ITF and remains unheated.
In the due course of an eventual resummation, however, consistency
may require to assign the auxiliary field a temperature dependent
Green function. This does, of course, not contradict the original
nature of an auxiliary field, since to a given order,
the perturbative inversion of
$\CD$ in the non-local Hamiltonian is not unique.

\section{The two-loop pressure}

In order to compare our method with the construction of James and Landshoff
\cit{jame1}, we calculate the two-loop pressure given by the diagrams depicted
in Fig1. We observe that apart from the third diagram, only transversal fields
contribute to the pressure.

\mkfig{
\begin{figure}
\hspace*{\fill}
\epsfbox{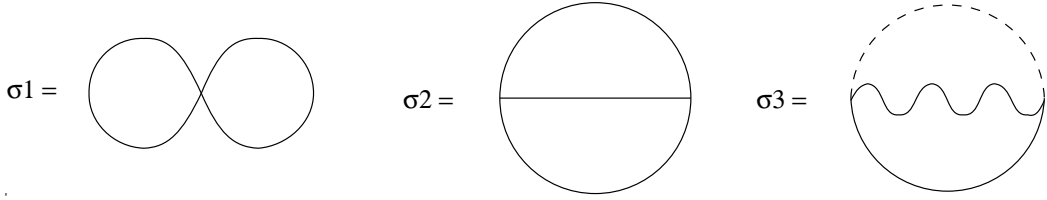}
\hspace*{\fill}
\vspace*{1em}
\caption{Graphs contributing to the two-loop pressure in the physical
 subspace. Full lines
correspond to $\hat A$ propagators, the wavy line to the $\hat E$-field
propagator and the broken line to the auxiliary field.}
\vspace{1em}
\end{figure}
}

The temperature dependent contribution reads  ($C_N=N (N^2-1) $ for SU(N) )
\beq  Z^{(2)} = -g^{2} C_{N}{ V \0 T }
\int{d^3p \0(2\pi )^{3}}\int{d^3q\0 (2\pi )^{3
}} \hat Z \lab{z2} , \eeq
where $\hat Z=  \sigma_1 + \sigma_2 + \sigma_3   $ and
$\sigma_i $ corresponds to the three graphs in Fig1 respectively,
$$
\sigma_1 =  {1\0 4}~{(3-z^{2})\0 \bp\bq} \( n_{p}n_{q} +n_{p} \),
\qquad
\sigma_2  =  {1 \0 2}~{z~(1+z^{2})\0 (\1p + \1q )^{2} }
\( n_{p} n_{q}+ n_{p} \) , $$
$$ \sigma_3  =  {1\0 4}~{1\0 \bp\bq}~{(1+z^{2})\0 (\1p +\1q
)^{2}}(\bp^{2}+\bq^{2})
\( n_{p}n_{q}+n_{p}  \), $$
$z = \vec p \vec q /{ p}{ q},\; p=|\1p | ,\;
n_p=(\exp ( \bp /T) -1 )^{-1} ,$
which gives the correct answer
$$ P^{(2)} = {T  \0  V} Z^{(2)} = -{g^{2}C_{N}T^{4} \0 144} . $$

Note the simplicity of the calculation. As opposed to that, the approach
of James and Landshoff involves
zero-temperature longitudinal propagators and
heated transversal ones. The calculation of the pressure calls for the
calculation of 12 graphs depicted in Fig2.

\mkfig{
\begin{figure}
\hspace*{\fill} \epsfbox{Fig2.eps} \hspace*{\fill}
\vspace*{1em}
\caption{Graphs contributing to the two-loop pressure in the approach of
James and Landshoff. The blob denotes an additional three-vertex due to
 the explicit construction of Gaus law states. }
\end{figure}
}

In addition to the
usual three- and four-vertices of QCD,
a static vertex denoted as a blob in Fig2 enters
from the explicit construction of physical states satisfying the Gauss
law constraint (\ref{gl}).  The (TTL) part of that vertex is
asymmetric in the transversal legs which we indicated by an asterisk.
We neither want to repeat the details of the formalism nor the lengthy but
straightforward calculation but rather state the result.
 Using the notation of Eq. (\ref{z2}) we find
$\hat Z = \sum_{i=0}^{10} \k_i $ where
 (The contributions
$\k_1 \ldots \k_4 $  have already been calculated in \cit{jame1}.)
$$  \k_{1}={1 \0 2}~{1\0 \bp\bq}~{(1+z^{2})\0 (\1p +\1q )^{2}}
\lb (n_{p}n_{q}+n_{p}) (\bp^{2}+\bq^{2}) +
(2n_{p}+1)  { 3 T\,  \bp^2 \bq  \0  \bp^2 + \bq^2 } \rb ,
 $$
$$ \k_{5}={p \0 8 T }~{(1-z^{2})\0 (\1p+\1q )^{2}}  (n_{p}+{1\0 2}) , $$
$ \k_0=\sigma_2,\;
\k_{2}=-{1 \0 2}\k_{1},\; \k_{3}=\k_{2},\; \k_4=\sigma_3 ,\;
\k_7=\k_{6}=\k_{5},
\; \k_{8}=-\k_{5},\; \k_{9}=0 ,\; \k_{10}=\sigma_1. $
All (TLL) graphs cancel. Furthermore, since $\k_1 + \k_2 + \k_3 = 0$
only the term $\k_4$ remains in the (TTL) contributions that coincides with
the third graph in Fig1 in our calculation.
It is interesting to observe that only the (TTL) part of
James and Landshoff's new vertex plays a r{\^o}le and that the only
non--vanishing contribution comes from diagrams which contain a
pair of that vertex part.

In the case of the two-loop
pressure, one may even replace the longitudinal $E$-fields in
the original Hamiltonian by the vertex part of the Gauss law constraint,
i.e. $ \1\9 \1E \to g f^{abc} E^{b,i} A^{c,i} $, and drop the
longitudinal $A$ fields to get the correct result. However, it is clear from
our investigation
that this oversimplifying guess does not give the correct
answer in general.

\section{Constructing Gauss law states to arbitrary order}

There are two different ways to handle gauge degrees of freedom
at finite temperature. Based on BRS invariance, Hata and Kugo \cit{hata}
constructed a theory where the Boltzman factor in the thermal average
gets replaced by $\exp{(i \pi N_c - H/T)}$, $N_c$ being the ghost number
operator, and the trace is expanded to include ghost fields and
all degrees of freedom of the gauge potential.
They demonstrated that with this weight thermal averages of operators
corresponding to observables are the same as
in the  projected ensemble. The advantage
of their approach, which is the standard way finite temperature
field theory is handled, is that all degrees of freedom are
heated which results in simple Feynman rules.
However, in the particular case of TAG, this construction does not work
\cit{nach1} and one has to go back to the projected ensemble involving
physical states only.

There, the thermal average of an observable $Q$ is defined by
$$ \li Q \re = Z^{-1} \sum \li \P \right|  e^{- H/T } Q \left| \P \re $$
where  physical states satisfy the Gauss law, $\CG
\left|\mbox{Phys} \re = 0 $. In the formalism of James and Landshoff those
are constructed by acting with  a unitary operator on the free transversal
states $\left| T \re$,
\beq  \left| \P \re = R \left|T \re ,\qd R = \sum_{n=0}^\infty g^n R_n
 \lab{22} \eeq
where $R_n$ was  determined by acting $n+1$ times with the Gauss law
operator on the physical states.
There the longitudinal components of the gauge potential correspond to
gauge degrees of freedom. In our approach,
the decomposition  (\ref{tra}) allows us  to identify the longitudinal
components with the 'angle' variable, $ A^L =  \9 \om $,
whereas the remaining field components read
\beq  A^T = A -  A^L = e^{ig\om} \hat A e^{-ig\om} -
{i \0 g } \left( \9 e^{ig\om} \right) e^{-ig\om} -  \9 \om. \lab{1} \eeq
Recalling that the field $\hat A$ contains but physical degrees of freedom,
one realizes that the transversal components of $A$ are physical
only  to zeroth order in $g$. Conversely, the choice to keep the longitudinal
components $A^L$ unheated to all orders corresponds to the fact, that the
physical field $\hat A$ is transversal only to lowest order, which in turn
means that there do appear higher order heated longitudinal modes contained
in the states that satisfy the Gauss law.

The operator $\hat A$
can be expressed in $(A^T,A^L)$ coordinates by virtue of  (\ref{1})
\beq \hat A =
\sum_{n=0}^\infty g^n  \hat A{}_n =
e^{-ig\om} \( A^T +( \9 \om )  +{i \0 g} \9 \) e^{ig\om} =
\sum_{n=0}^\infty { (ig)^n \0 n! } \stackrel{(n)}{ [ } \om , A^T + {n \0 n+1}
 \9 \om ]  \eeq
where
$\stackrel{(n)}{[} X,Y] = [X,\stackrel{(n-1)}{[} X,Y] ],\,
\stackrel{(0)}{[} X,Y]=Y$ denotes
the multiple commutator.
$\hat A$ is the counterpart of the Gauss law operator to arbitrary order,
where  the first few terms read
$$ \hat A = A^T  + i g [\om , A^T + {1 \0 2} \9 \om ] - {g^2 \0 2}
[\om , [ \om , A^T + { 2 \0 3 } \9 \om ]] + \ldots .$$

Since the change of the basis from physical to transversal states
is mediated by a unitary transformation, the corresponding operators
$\hat A$ and $A^T$ are unitary equivalent according to
${\hat A} R = R A^T .$
Collecting terms by orders of $g$, this leads to the recursion
 relation
$$  [\hat A_0 ,R_n] + \sum_{m=0}^{n -1}  \hat A{}_{n-m} R_m  = 0 . $$
The strategy to solve this recursion may be
motivated by the following observation.
$R_n$ is given by the action of an (unknown) operator on the sum
which inverts the commutator with  $\hat A_0=A^T$. Recalling
the equal time commutator
$[A^{Ti,a}(x),E^{Tj,b}(y)] = i \de^{ab} T^{ij}(\9) \de (x-y) $,
one may, roughly speaking, construct $R_n$ by 'multiplying' the sum with $E^T$.

In particular, for $n=1$ one has to study the equation
$  [ {\hat A}_0,R_1 ] + {\hat A}_1  =0 $
that has the solution
$$ R_1 = i \; ( E^T \cdot \hat A_1 ) :=
i \int d^3z \1E{}^{T,a}(z) {\hat {\1 A_1^a}}(z) .$$
Plugging in the expression for $\hat A_1$, the explicit
form of $R_1$ can be written as ($\om = \9A^L / \9^2 $)
$$ R_1 = i f^{abc} \int d^3 z \( {1 \0 \9^2 } \9 A^{L,a} \)\! (z) \, \1E^{T,b}
\( \1A^{T,c} (z) + {1 \0 2 } \1A^{L,c} (z) \)  $$
which coincides with the result of James and Landshoff.

The recursion for $n=2$ reads
\beq  [ {\hat A}_0,R_2 ] + {\hat A}_1 R_1 + {\hat A}_2  = 0 .\lab{rec2}
\eeq
Guided by what we have learned above, one would naively guess
$R^{guess}_2 = i \; ( E^T \cdot \hat A_2 ) + \half R_1  R_1  $ which when
 commutated with
$A_0$ gives $ [ {\hat A}_0 , R^{guess}_2 ] = - {\hat A}_2 - \half
{\hat A}_1 R_1  -
\half R_1 {\hat A}_1  $ that  cancels the third but not the
 second term in the
recursion (\ref{rec2}) since  $\hat A_1 $ does not commute with $R_1$.
We therefore add a suitable chosen term proportional to the commutator
which compensates the wrong order
in the $R_1 \hat A_1$ contribution. The solution of (\ref{rec2}) reads
$$ R_2 = i \, (E^T \cdot \ah_2 ) + {1 \0 2} R_1 R_1 - {i \0 2 } ( E^T \cdot
[R_1 , \hat A_1] ) $$
which is unique up to operators that commute with $\hat A_0$.

One may continue further and calculate the $n=3$ contribution to $R$,
\begin{eqnarray*}  R_3 = {1 \0 6}R_1^3 +{i \0 3} \(\et \cdot
\stackrel{(2)}{[}R_1,\ah_1] \) -
{i \0 2} R_1 \(\et \cdot [R_1,\ah_1] \) +
 i R_1 \(\et \cdot \ah_2 \) - \\ -  i \(
\et \cdot [R_1,\ah_2 ]\) +  i \( \et \cdot \ah_3 \) , \end{eqnarray*}
and a calculation of higher order terms proceeds analog
similar lines. We note that our construction has formal similarity with
the Foldy-Wouthuysen transformation in quantum-mechanics.

Our result does not match
the $R_2$ contributions found in \cit{wong} that also contain
time derivatives of the longitudinal fields, but agrees with the
argument given by James and Landshoff that those time derivatives should not be
present in $R$.
We also find that the exponentiation conjectured in
\cit{wong} cannot be confirmed. Although the $R_1$ terms appear with
the correct factors, a complete exponentiation is spoiled by an
increasing number of commutator terms, which is consistent with the
nonlocal Hamiltonian (\ref{ham}) that also contains an infinite number of
vertices. Unlike as in pure QED, where the radiation gauge
eliminates all gauge freedoms from the Hamiltonian in a local manner, and
where $R$ does  exponentiate,
is has been argued \cit{creu} that in nonabelian gauge theory no such
canonical gauge exists.

\section{Conclusion and Outlook}

We formulated a finite temperature framework for pure QCD
which is based on the elimination of all gauge degrees of freedom.
In contrast to the former approach of James and Landshoff,
who explicitly constructed physical states by
solving the Gauss law constraint, we eliminate spurious
degrees of freedom  at the operator level which allows to heat
the remaining degrees of freedom in a
straight-forward way. We do not encounter
any pole ambiguities which exist in the naive imaginary time formulation of
axial gauges.

Although  our effective  Hamiltonian contains a
 kinetic term non-local in the fields, it is possible to find a local
formulation by introducing an  auxiliary field.
We compared our theory with the construction of
James and Landshoff for the particular case of the two-loop pressure and
found that the number of Feynman graphs is reduced drastically in our
framework. Since our construction allows to make a clear distinction
between physical and gauge degrees of freedom, we can also
line out a strategy of how to explicitly construct Gauss law states to all
orders. Those are calculated explicitly up to third order in  the
basis of the free transversal states of the gauge field.

Clearly, we will  establish a resummation
program in our formalism. Since the Hamiltonian only contains
physical observables, the location of the poles in the propagators
 contains intrinsic physical information on the
the on-shell
dispersion relation. It would be interesting to compare with
results obtained in the usual approach, where gauge-fixing
independence of the poles has to be and was proven \cit{rebh3}.

Furthermore, we only dealt with pure QCD, which by construction excludes
the calculation of the Debye-mass that would
require the gauge-invariant inclusion of a charge
density in the Gauss law constraint. Conversely, if it turned out
to be  possible to include charges in the present formalism it would
be possible to separate effects from external charges and those
induced by pure QCD. We are going to investigate on this question.


\begin{references}
\bibit{grib1} V.~N.~Gribov, Nucl.~Phys.\ { B 139} (1978) 1;
I.~M.~Singer, Commun.~Math.~Phys.\ { 60} (1978) 7;
M.~A.~Soloviev, Theor.~Math.~Phys.\ (USSR) {78} (1989) 117.
\bibit{LM} S.~Mandelstam, Nucl.~Phys.\ B 213 {(1983) 149};
G.~Leibbrandt, Phys.~Rev.\ D 29 {(1984) 1699};
Phys.~Rev.\ D 30 {(1984) 2167};
G.~Leibbrandt and S.~L.~Nyeo, Phys.~Rev.\ D {(1986) 3135}.
\bibit{gaig1} P.~Gaigg and M.~Kreuzer, Phys.~Lett.\ B 205 (1988) 530.
A general reference on nonstandard gauges is:
P.~Gaigg, W.~Kummer and M.~Schweda, eds. {\em
Physical and nonstandard gauges}, Lecture Notes in Physics, Vol.\ 361
(Springer, Berlin, 1990).
\bibit{land0} P.~V.~L.~Landshoff, Phys.~Lett.\ B 169 (1986) 69.
\bibit{nach1} M.~Kreuzer and H.~Nachbagauer, Phys.~Lett.\ {B 271} (1991) 155.
\bibit{jame0} K.~A.~James, Z.~Phys.\ {C 48} (1990) 169; Z.~Phys.\ {C 49} (1991)
115.
\bibit{nach2} H.~Nachbagauer, Z.~Phys.\ {C 56} (1992) 407.
\bibit{kaja1} K.~Kajantie and J.Kapusta, Ann.~Phys.~ (N.Y.) 169 (1985) 477;
U.~Heinz, K.~Kajantie and T.~Toimela, Ann.~Phys.~(N.Y.) 176 (1987)
218;
J.~Kapusta, {\em Finite temperature field theory}, Cambridge University
Press, 1989.
\bibit{resum} E.~Braaten and R.~D.~Pisarski,  Nucl.~Phys.\ { B 337} (1990)
569.
\bibit{peig1} S.~Peign\'e and S.~M.~H.~Wong,
Non-Debye screening in finite temperature QCD,
preprint LPTHE-Orsay 94/46 (1994), hep-ph 9406276.
\bibit{baie1} R.~Baier and O.~K.~Kalashnikov, Phys.~Lett.\ {B 328} (1994)
450.
\bibit{rebh2} A.~K.~Rebhan, Nucl.~Phys.\ { B 430}(1994) 319.
\bibit{braa1} E.~Braaten and A.~Nieto,  Phys.~Rev.~Lett.\ 73 (1994) 2402.
\bibit{rebh1} A.~K.~Rebhan, Phys.~Rev.\ { D 48 }(1993) R3967.
\bibit{kala1} O.~K.~Kalashnikov, The High Temperature Dispersion
Equation for Longitudinal Plasma Oscillations in TAG,
preprint BI-TP 95/06, hep-ph 9502354.
\bibit{jame1} K.~James and P.~V.~Landshoff, Phys.~Lett.\ {B  251} (1990) 167.
\bibit{shab1} S.~V.~Shabanov, 2-D Yang-Mills Theories, Gauge Orbit
Spaces and the Path Integral Quantization, preprint SACLAY-SPHT-93-139,
hep-th 9312160.
\bibit{shab2} S.~V.~Shabanov, Phys.~Lett.\ { B 255} (1991) 398.
\bibit{dira} P.~A.~M.~Dirac, Lectures in quantum mechanics, (Yeshiva
University,
New York, 1964); A.~Hansson, T.~Regge and C.~Teitelboim,
{\em Constrained Hamiltonian Systems}, Accademia Nationale dei Lincei (1976).
\bibit{creu} M.~Creutz, I.~J.~Muzinich and T.~N.~Turdon, Phys.~Rev.\ { D
19} (1979) 531.
\bibit{turd} T.~N.~Turdon, Phys.~Rev.\ {D 21} (1980) 2348.
\bibit{wong}  S.~M.~H.~Wong,  Preliminary study of a new formalism of
  temporal axial gauge at finite~T, preprint LPTHE-Orsay 94/114, hep-th
9501032.
\bibit{hata} H.~Hata and T.~Kugo, Phys.~Rev.\ D 21 (1980) 3333;
 T.~Kugo and I.~Ojima, Progr.~Theor.~Phys.~Suppl.\ 66 (1980) 1.
\bibit{rebh3} R.~Kobes, G.~Kunstatter and A.~Rebhan, Nucl.~Phys.\
 B 355 (1991) 1.


\end{references}
\end{document}